\documentclass[9pt, sigconf]{acmart}

\AtBeginDocument{%
  }
 
\setcopyright{none}
\acmDOI{}
\acmISBN{}
\acmBooktitle{}
\acmConference[This paper is to be presented in DAC '26]{Design Automation Conference}{July 26--29, 2026}{Long Beach, CA}

\usepackage{pgfplots}
\usepackage{algorithm}
\usepackage{algorithmic}
\usepackage{graphicx}
\usepackage{textcomp}
\usepackage{xcolor}

\usepackage{booktabs} 
\usepackage[caption=false]{subfig}
\usepackage{fancyvrb} 
\usepackage{multirow}
\usepackage{diagbox}
\usepackage{tikz}
\usepackage{numprint}
\usepackage{url}
\usepackage{colortbl}
\usepackage{dhucs} 

\usepackage{enumitem} 
\usepackage{makecell} 

\usepackage[font=small]{caption}
\usepackage[font=footnotesize]{subcaption}
\usepackage{comment}
\usepackage{lineno}
\usepackage{color} 
\usepackage{setspace} 
\usepackage{relsize}
\usepackage{tabularx}
\usepackage{bm}
\usepackage{hhline}
\usepackage{xurl}

\usepackage{ragged2e}
\usepackage{stfloats}

\pgfplotsset{compat=1.18}
\setcitestyle{numbers,sort&compress}

\newcommand\refFigure[1]{Figure~\ref{#1}}
\newcommand\refTable[1]{Table~\ref{#1}}
\newcommand\refSection[1]{Section~\ref{#1}}
\newcommand\refAlgo[1]{Algorithm~\ref{#1}}
\newcommand\refEqn[1]{(\ref{#1})}

\newcommand\korean[1]{}
\newcommand\blind[1]{XXXX}

\renewcommand{\baselinestretch}{1}

\begin{document}
\title[TT-SEAL: TTD-Aware Selective Encryption for Adversarially-Robust and Low-Latency Edge AI]{TT-SEAL: TTD-Aware Selective Encryption for Adversarially-\\Robust and Low-Latency Edge AI}

\author{Kyeongpil Min, Sangmin Jeon, Jae-Jin Lee, and Woojoo Lee
}
\affiliation{%
  \institution{\vspace{0.1em} $^{ \dag}$Department of Intelligent Semiconductor Engineering, Chung-Ang University, Seoul, Korea \\ \vspace{0.1em}
$\ddag$ AI Edge SoC Research Section, Electronics and Telecommunications Research Institute, Daejeon, Korea \\}
  \country{} 
}
\email{}

%
%

\thanks{
This work was supported in part by Institute of Information \& communications Technology Planning \& Evaluation (IITP) grants funded by the Korea government (MSIT) (No. RS-2023-00277060), and in part by the National Research Foundation of Korea (NRF) grant funded by MSIT (No. RS-2024-00345668). 

Kyeongpil Min and Sangmin Jeon contributed equally to this work. 

Woojoo Lee is the corresponding author.
}

\begin{abstract}

Cloud--edge AI must jointly satisfy model compression and security under tight device budgets. While Tensor-Train Decomposition (TTD) shrinks on-device models, prior selective-encryption studies largely assume dense weights, leaving its practicality under TTD compression unclear. We present \emph{TT-SEAL}, a selective-encryption framework for TT-decomposed networks. TT-SEAL ranks TT cores with a sensitivity-based importance metric, calibrates a one-time robustness threshold, and uses a value-DP optimizer to encrypt the minimum set of critical cores with AES. Under TTD-aware, transfer-based threat models---and on an FPGA-prototyped edge processor---TT-SEAL matches the robustness of full (black-box) encryption while encrypting as little as 4.89--15.92\% of parameters across ResNet-18, MobileNetV2, and VGG-16, and drives the share of AES decryption in end-to-end latency to low single digits (e.g., 58\%\,$\rightarrow$\,2.76\% on ResNet-18), enabling secure, low-latency edge AI.

\end{abstract}

\maketitle

\section{Introduction}\label{sec:intro}

Cloud-edge collaborative AI leverages cloud resources to train large-scale neural networks while enabling low-latency inference on edge devices~\cite{wang2023shoggoth, zhang2024edge}. 
This architecture has accelerated the deployment of intelligent services such as speech recognition, video analytics, and IoT applications by flexibly distributing workloads between cloud and edge. 
However, frequent model updates require transmitting large parameter sets, and edge devices—operating under tight storage, computation, and energy budgets—face persistent challenges in efficiently storing and synchronizing models~\cite{shuvo2022efficient, wang2022edcompress}. 

To address these challenges, numerous model compression techniques have been studied, including quantization~\cite{shen2025quartdepth}, pruning~\cite{he2023structured}, knowledge distillation~\cite{suwannaphong2025optimising}, and low-rank or tensor factorization~\cite{liu2023tensor}. 
Among these, Tensor-Train Decomposition (TTD)~\cite{oseledets2011tensor} has shown particular promise by decomposing a high-dimensional weight tensor into a sequence of low-dimensional cores. 
TTD preserves the structural properties of layers while reducing redundancy, yielding models with much lower parameter counts and storage complexity. 
As a result, it reduces model size, communication volume, and memory-access cost during inference, thereby enabling efficient deployment under edge constraints~\cite{Kwak:DATE2025,dong2025duqtta, Kwak:Math2025}. 

Beyond efficiency, however, cloud–edge collaborative AI raises significant \emph{security} concerns. 
Unlike conventional cloud-only models where parameters remain protected in datacenters, distributed deployment requires storing parameters in off-chip DRAM and transmitting them over external buses on edge devices, exposing them to physical attacks~\cite{gai2021attacking, wang2023security}. 
Compromised weights can lead to IP theft, reverse engineering~\cite{hua2018reverse}, model cloning~\cite{kariyappa2020defending}, and adversarial attacks~\cite{kurakin2018adversarial, yuan2019adversarial}, ultimately threatening the reliability of IoT devices~\cite{bao2021threat, khowaja2022get} and autonomous systems~\cite{qayyum2020securing}. 
Notably, when applying adversarial transferability attacks~\cite{papernot2017practical, goodfellow2018making} to TTD-compressed models, we observe that despite reduced representational capacity and constrained decision boundaries, substitute models can still approximate the original model’s behavior closely. 
This finding indicates that TTD compression alone does not mitigate adversarial transferability risks, and adversarial threats remain a pressing security concern.

A straightforward defense is to encrypt all model parameters before storing or transmitting them. 
While full encryption provides strong protection, it imposes significant latency and energy overhead because every inference requires heavy decryption at the edge. 
To reduce this overhead, \emph{selective encryption} has been explored. 
Instead of protecting all weights, selective encryption targets only a subset of parameters, approximating the robustness of full encryption at a fraction of the cost~\cite{zuo2021sealing, Tian2021Probabilistic, hu2024maskcrypt}. 
Examples include L1-norm–based selection of convolutional kernel rows~\cite{zuo2021sealing}, probabilistic masking~\cite{Tian2021Probabilistic}, encrypting only a portion of gradient updates in federated learning~\cite{hu2024maskcrypt}, or combining partial functional encryption with adversarial training~\cite{ryffel2019partially}. 
These works collectively show that encrypting only part of a dense model can maintain robustness against inference attacks while greatly reducing cost. 

However, nearly all prior approaches assume \emph{dense} weight representations. 
Directly applying such dense-oriented selective encryption to TTD-compressed models is problematic. 
In dense networks, redundancy allows partial encryption (e.g., 50\% of weights) to achieve robustness close to full encryption. 
In contrast, TTD removes much of this redundancy, concentrating information into fewer parameters; leaving even a small portion unencrypted can still expose critical structure. 
This fundamental difference limits the effectiveness of existing methods and calls for a TTD-aware selective encryption scheme.

In this work, we present \textbf{TT-SEAL}—\emph{Selective Encryption for Adversarially-Robust and Low-Latency}—to our knowledge, the first selective encryption framework explicitly tailored to \emph{TTD}-\hspace{0pt}compressed neural networks. 
TT-SEAL leverages the structural properties of TT-format weights to achieve strong protection while minimizing encryption overhead. 
Concretely, TT-SEAL introduces (i) a core-wise importance metric that evaluates the security impact of each TT-core, (ii) a data-driven threshold calibration that aligns protection strength with a robustness target, and (iii) a minimal-cost selection algorithm that efficiently identifies the smallest set of TT-cores to encrypt using dynamic programming-based optimization. 
Through extensive experiments, we demonstrate that TT-SEAL significantly reduces decryption overhead while maintaining robustness comparable to full encryption, thereby enabling secure and low-latency inference on edge devices.

The contributions of this work are as follows:

\begin{itemize}[leftmargin=*,nosep]
\item \textbf{TTD-tailored selective encryption.} We introduce TT-SEAL, the first selective encryption method specifically designed for TTD-compressed models, exploiting TT-core structure for strong protection with reduced overhead.  
\item \textbf{Core-level metric and optimization.} We define a sensitivity-based importance metric, calibrate a robustness threshold, and formulate an optimization procedure to identify the minimal set of TT-cores to encrypt.
\item \textbf{Prototype-based validation.} On an FPGA-based edge AI processor, TT-SEAL achieves robustness comparable to full encryption while encrypting as little as 4.89\% of parameters (ResNet-18), with similar trends observed on VGG-16 and MobileNetV2. Moreover, the decryption share in end-to-end inference is reduced from 58\% to 2.76\%, making secure edge inference feasible.
\end{itemize}

\begin{figure}[t]
\centering
\includegraphics[width=0.97\columnwidth]{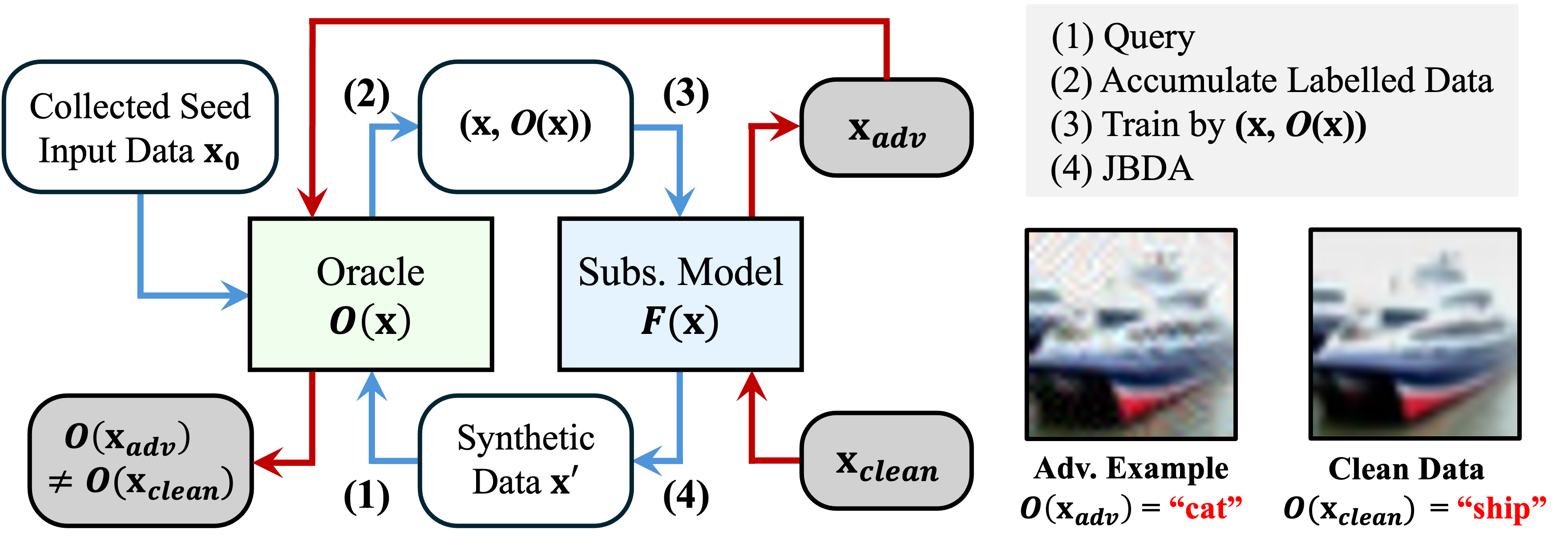}
\vskip -6pt
\caption{Transfer-based adversarial attack using JBDA.  
Blue lines: the attacker queries oracle $\boldsymbol{O(x)}$ with clean inputs, trains a substitute $\boldsymbol{F(\mathbf{x})}$, and augments data $\mathbf{x}'$ near decision boundaries.  
Red lines: $\boldsymbol{F}$ generates adversarial examples $\mathbf{x}_{\boldsymbol{adv}}$ that transfer to $\boldsymbol{O}$.}
\label{fig:JBDA}
\vskip -4pt
\end{figure}

\section{System Context and TTD-Aware Threat Model}\label{Motivation}

\subsection{TTD Compression for Deployed DNNs} \label{sec:TTD}
TTD represents a $d$-dimensional weight tensor $\mathcal{W}$ as the product of $d$ low-dimensional core tensors $G_k$~\cite{oseledets2011tensor}:
\begin{align} \label{eqn:TTD}
	\begin{gathered}
		\mathcal{W}(i_1,i_2,\cdots,i_d) \approx G_1[i_1] \, G_2[i_2] \cdots G_d[i_d], \\
		\mathcal{W} \in \mathbb{R}^{n_1 \times n_2 \times \cdots \times n_d}, \quad G_k \in \mathbb{R}^{r_{k-1} \times n_k \times r_k},
	\end{gathered}
\end{align}
where $n_k$ denotes the size of the $k$-th mode and $r_k$ the TT-rank with $r_0=r_d=1$.  
This TT-format reduces parameter count and storage cost while preserving the layer structure.  

Applied to neural networks, TTD compresses convolutional, recurrent, and transformer models with minimum accuracy loss. For example, a convolutional layer has a four-dimensional weight tensor $\mathcal{W}_{conv} \in \mathbb{R}^{C_{out}\times C_{in}\times K_h\times K_w}$, where $C_{in}$ and $C_{out}$ denote the number of input and output channels, and $K_h$ and $K_w$ represent the kernel height and width, respectively. The storage complexity of $\mathcal{W}_{conv}$ is $O(C_{out} C_{in} K_h K_w)$, while its TT-format counterpart requires only $O(r^2(C_{out}+C_{in}+K_h+K_w))$. 
This reduction alleviates memory footprint, reduces transmission volume, and lowers inference cost, which is particularly beneficial for edge devices~\cite{lee2024tt, wei2024fmtt, han2023multiscale}.

\subsection{TTD-Aware Threat Model for Edge AI}\label{sec:TTD-threat}

\textbf{System setting.} An edge SoC integrates compute units (CPUs or accelerator cores), system interconnect, on-chip memory (registers, caches, eSRAM), and off-chip DRAM. Large neural network weights cannot fit on-chip and are stored in compressed form in DRAM. During inference, these parameters are fetched over the DDR bus, decompressed on-chip, and then loaded into compute units for execution.

\textbf{Trust boundary.} The on-chip region is physically embedded and thus difficult to probe directly, whereas off-chip DRAM and memory buses are external and exposed to adversaries. 
We therefore assume model parameters in DRAM are always encrypted as a baseline protection. 
Nevertheless, outputs (e.g., labels, logits, or scores) remain observable through normal I/O interfaces and can be collected by an attacker.

\textbf{Adversary capability.} We assume (i) encrypted parameters in DRAM can be observed by the attacker, and (ii) the attacker can repeatedly query the deployed model (oracle $O$) to collect input-output pairs $(\mathbf{x}, O(\mathbf{x}))$. Even when all parameters remain encrypted (black-box setting), such queries allow the attacker to train a substitute model $F$ that approximates $O$.  

\textbf{Attack procedure.} \refFigure{fig:JBDA} illustrates the process: 
in the \emph{blue path}, the attacker queries the oracle $O(x)$ with clean inputs, 
trains a substitute $F(x)$, and augments data near decision boundaries using Jacobian-based Dataset Augmentation (JBDA)~\cite{papernot2017practical}. 
In the \emph{red path}, the trained $F$ generates adversarial examples $\mathbf{x}_{adv}$ that transfer to $O$. 
A common generator is the Fast Gradient Sign Method (FGSM)~\cite{goodfellow2014explaining, kurakin2018adversarial}:
\begin{align}\label{eqn:FGSM}
\mathbf{x}_{adv} = \mathbf{x} + \epsilon \cdot \text{sign}\!\left(\nabla_{\mathbf{x}} L(\mathbf{x}, y_{\text{true}})\right),
\end{align}
where $y_{\text{true}}$ is its ground-truth label, $L(\cdot,\cdot)$ is the loss (e.g., cross-entropy), $\nabla_{\mathbf{x}}$ is the gradient w.r.t.\ the input, and $\operatorname{sign}(\cdot)$ is applied element-wise. 
Larger $\epsilon$ values typically raise the chance of misclassification but also make perturbations more visible, whereas smaller values keep inputs visually closer to the original but reduce attack success. 
Both non-targeted (any misclassification) and targeted (forcing a specific label) variants are considered realistic threats in this work.

\textbf{Implication for compressed models.} The effectiveness of such transfer-based attacks depends critically on how well $F$ approximates $O$: the closer the approximation, the more likely adversarial examples crafted on $F$ will transfer to $O$. In practice, full secrecy of weights constrains $F$, but even partial parameter exposure can substantially improve its accuracy and thus boost attack transferability~\cite{goodfellow2018making}. Our experiments confirm that TTD compression does not eliminate this risk: on CIFAR-10, substitute models trained in a white-box setting achieved accuracies of 92.29\% (ResNet-18~\cite{he2016deep}), 89.5\% (MobileNetV2~\cite{sandler2018mobilenetv2}), and 90.73\% (VGG-16~\cite{simonyan2014very}). Despite parameter reduction and redundancy removal, TTD models still leak sufficient structural information for substitutes to approximate decision boundaries, leaving them vulnerable to transfer-based adversarial attacks.

\subsection{Limits of Selective Encryption in TT Models}\label{sec:limitations}

\begin{figure}[t]
\centering
\includegraphics[width=0.95\columnwidth]{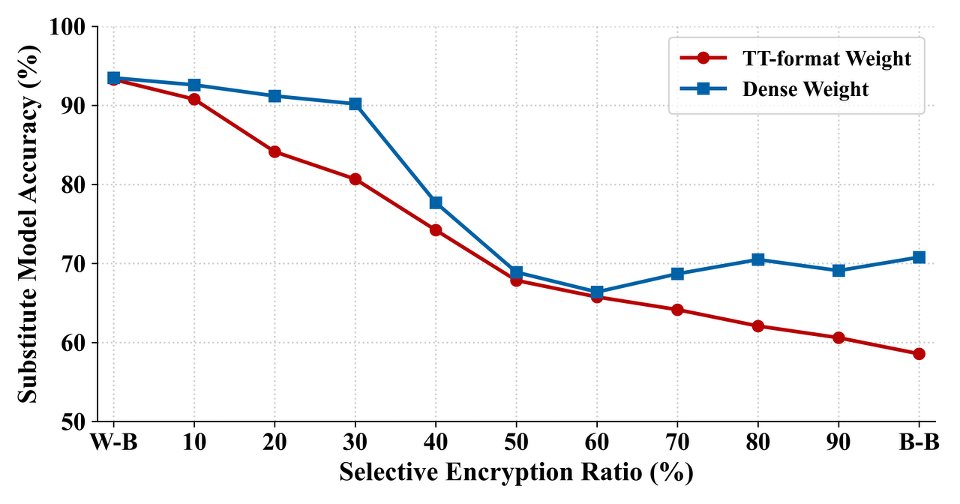}
\vskip -10pt
\caption{Substitute-model accuracy vs.\ selective encryption ratio in ResNet-18 with dense and TTD-compressed weights.}
\label{fig:SE_base_TTD}
\vskip -4pt
\end{figure}

Selective encryption seeks to reduce the overhead of full encryption by protecting only a subset of parameters while preserving robustness. This approach is attractive for edge deployment where full encryption incurs excessive latency and power.  

Prior work has explored several strategies: kernel-row selection based on L1 norm (showing 50\% suffices for dense CNNs)~\cite{zuo2021sealing}, probabilistic masking (8\% encrypted)~\cite{Tian2021Probabilistic}, selective gradient encryption in federated learning (up to $4.15\times$ communication savings)~\cite{hu2024maskcrypt}, and partial functional encryption combined with adversarial training~\cite{ryffel2019partially}.  
These results collectively show that encrypting only part of a model can be both feasible and effective---yet nearly all assume dense weight representations.

To examine the TT setting, we applied~\cite{zuo2021sealing} to both dense and TTD-compressed ResNet-18 on CIFAR-10.  
\refFigure{fig:SE_base_TTD} plots substitute accuracy versus encryption ratio. Using the unencrypted model as the \emph{white-box} (W-B) reference and the fully encrypted model as the \emph{black-box} (B-B) baseline: the dense model (blue) reached B-B robustness with only $\sim$50\% encrypted weights, whereas the TTD model (red) required encrypting $>90\%$ yet still failed to match B-B.  

This discrepancy stems from information density. Dense models contain redundancy, so unencrypted portions often hold less critical information. TT-format removes redundancy during decomposition; remaining parameters are more concentrated and thus reveal more if left unencrypted. Consequently, directly applying dense-oriented selective encryption to TT-format models is ineffective.  
These observations motivate a selective encryption scheme specifically tailored to the TT structure, which we develop in \refSection{sec:method}.

\section{TT-SEAL: Proposed Method}\label{sec:method}
This section presents TT-SEAL, a selective encryption framework tailored to {TTD}-compressed neural networks. 
As discussed in \refSection{sec:limitations}, TT-format weights differ from dense representations in two key ways: 
redundancy is largely removed by decomposition, and information is structurally distributed across sequential TT-cores as defined in \refEqn{eqn:TTD}. 
These properties weaken dense-oriented selection rules, but they also imply that encrypting only a subset of cores can significantly obfuscate the original weights. 
TT-SEAL leverages this structure through three components: a core-wise importance metric, a data-driven security threshold, and a minimal-cost selection algorithm.

\begin{algorithm}[t]
\footnotesize
\caption{Computation of $I_{acc}$ for TT-cores}\label{alg:iacc}
\begin{algorithmic}[1]
\STATE \textbf{Input:} TTD-compressed model $M$, validation set $\mathcal{D}_{val}$, \# of probes $N_v$
\STATE For each core $G_{l,k}$, initialize accumulators for gradient norms.
\FOR{mini-batch $\mathcal{B}\subset\mathcal{D}_{val}$}
  \STATE Forward pass to obtain $y$ and loss $L$.
  \STATE Backward pass on $L$ to accumulate $\|\partial L/\partial G_{l,k}\|_F^2$.
  \FOR{$t=1$ to $N_v$}
    \STATE Sample $\mathbf{v}\sim\mathcal{R}$ and compute gradient of $s(\mathbf{v})=\mathbf{v}^\top y$.
    \STATE Accumulate $\|\partial s(\mathbf{v})/\partial G_{l,k}\|_F^2$.
  \ENDFOR
\ENDFOR
\STATE Normalize by $\mu_l=\mathrm{mean}_k \|G_{l,k}\|_F$ per layer.
\STATE Return $I_{acc}(G_{l,k})$ for all cores.
\end{algorithmic}
\end{algorithm}

\noindent \textit{\textbf{Problem Setup and Security Target}\label{sec:target}}
\\ \indent 
Let $\{G_{l,k}\}$ denote the set of TT-cores across all layers $l$, with $k$ indexing the cores in layer $l$, and let $size(G_{l,k})$ be the number of parameters in core $G_{l,k}$. 
Following the threat model in \refSection{sec:TTD-threat}, the \emph{security target} is to identify the smallest set $S \subseteq \{G_{l,k}\}$ to encrypt such that the accuracy of a substitute model does not exceed the fully encrypted black-box baseline by more than $\delta$ (we use $\delta=3\%$). 

Let $A_{\mathrm{BB}}$ denote the baseline accuracy under full black-box encryption. 
TT-SEAL enforces the constraint
\begin{align}
A_{\mathrm{sub}}(S) \;\le\; A_{\mathrm{BB}} + \delta,
\end{align}
where $A_{\mathrm{sub}}(S)$ is the accuracy of a substitute model given encrypted set $S$, while minimizing the total encryption cost
\begin{align}
C_{\mathrm{enc}} = \sum_{G \in S} size(G).
\end{align}

\noindent \textit{\textbf{Core-wise Importance Metric}\label{sec:metric}}
\\ \indent 
To score how critical each core is to accuracy and decision boundaries, we define the \textbf{Importance Metric} $I_{acc}$:
\begingroup
\fontsize{9pt}{11pt}\selectfont
\begin{align}\label{equ:app}
	I_{acc}(G_{l,k}) = \frac{1}{\mu_{l}}
	\sqrt{\mathbb{E}_{\mathbf{x}\in\mathcal{D}_{val}}\!\left[\left\|\frac{\partial L}{\partial G_{l,k}}\right\|_F^{2}
	  + \left\|\frac{\partial y}{\partial G_{l,k}}\right\|_F^{2}\right]},
\end{align}
\endgroup
where $L$ is the training loss, $y$ the model output (post-softmax vector), $\|\cdot\|_F$ the Frobenius norm, $\mathcal{D}_{val}$ a held-out set, and $\mu_l$ the average Frobenius norm of all cores in layer $l$ (for cross-layer normalization). Intuitively, $I_{acc}$ aggregates first-order sensitivities of both output and loss w.r.t.\ a core; a larger value implies that encrypting this core is more likely to suppress substitute accuracy.

\textit{a) Practical estimation.:}
The term $\partial L/\partial G$ is obtained from a standard backward pass. The Jacobian norm $\|\partial y/\partial G\|_F$ can be estimated efficiently via a Hutchinson-type trick:
\begin{align}\label{equ:Hutchinson}
\big\|\tfrac{\partial y}{\partial G}\big\|_F^2
~=~ \mathbb{E}_{\mathbf{v}\sim \mathcal{R}}\!\left[\left\|\tfrac{\partial (\mathbf{v}^{\top} y)}{\partial G}\right\|_F^2\right],
\end{align}
where $\mathcal{R}$ is a Rademacher or standard normal distribution over the output dimension. In practice, one computes gradients of the scalar $\mathbf{v}^\top y$ for $N_v$ random vectors ($N_v\!\in\![1,4]$ typically suffices), averages their squared Frobenius norms across a small validation subset, and plugs the estimate into \refEqn{equ:app}. The full computation procedure is summarized in \refAlgo{alg:iacc}.

\textit{b) Normalization and robustness.:} We use $\mu_l = \mathrm{mean}_{k}\,\|G_{l,k}\|_F$ for scale invariance across layers; 
median or percentile-based options are also viable to dampen outliers. 
Mini-batch estimates over $\mathcal{D}_{val}$ are averaged to mitigate stochasticity.

\begin{figure}[t]
\centering
\includegraphics[width=0.95\columnwidth]{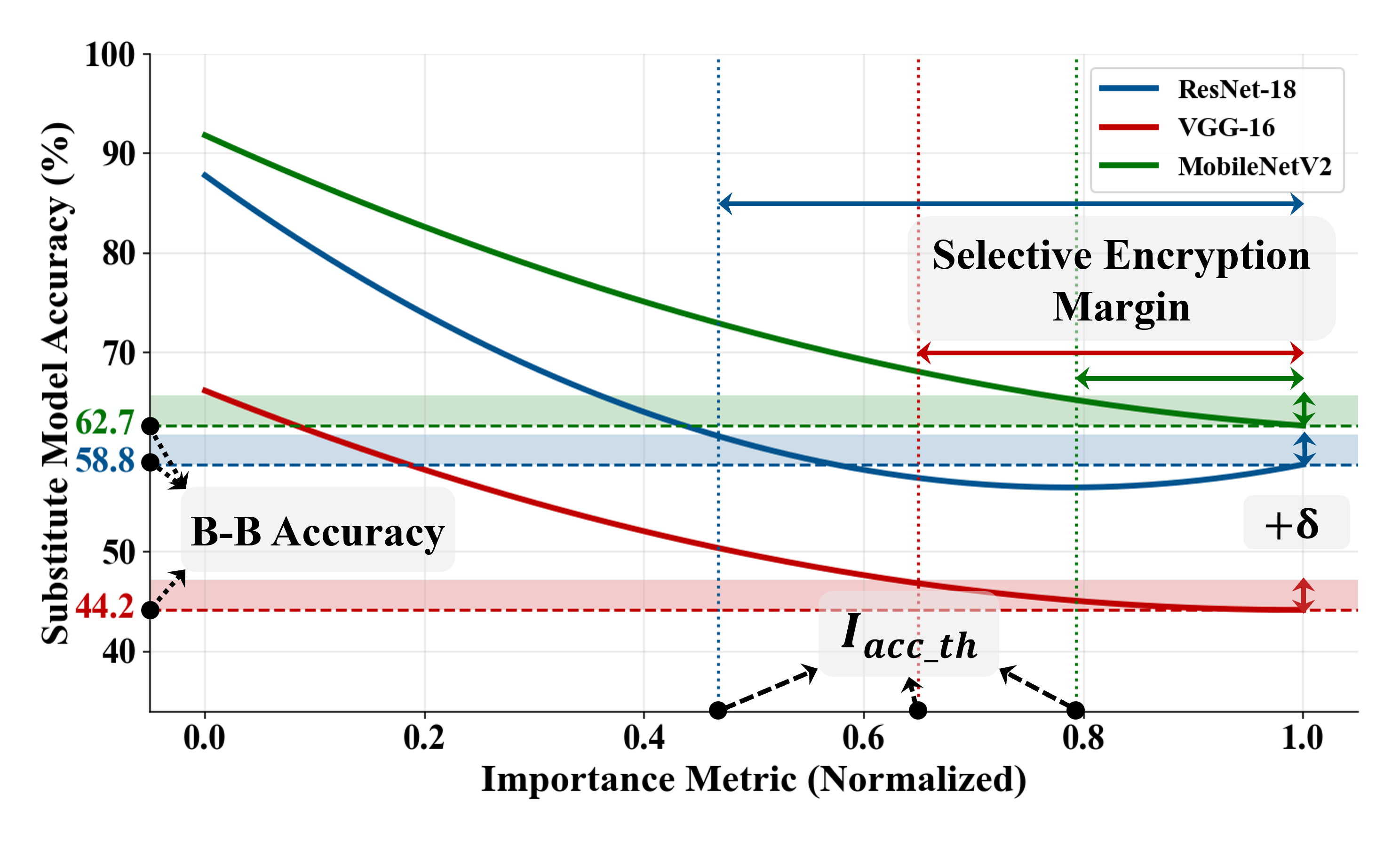}
\vskip -10pt
\caption{Relationship between $\boldsymbol{I}_{\boldsymbol{acc}}$ and substitute-model accuracy for TTD-compressed ResNet-18, MobileNetV2 and VGG-16.}
\label{fig:I_acc}
\vskip -4pt
\end{figure}

\textit{c) Empirical monotonicity.:}
As illustrated in \refFigure{fig:I_acc}, encrypting cores in ascending $I_{acc}$ order yields a monotone decrease in substitute accuracy that approaches the B-B baseline. 
Occasional local rebounds may appear due to overlapping information across TT-cores and stochastic factors in substitute training (e.g., mini-batch sampling, random initialization, SGD noise, core-combination effects), but the overall trend decreases and converges to the B-B robustness level.

\noindent \textit{\textbf{Threshold Calibration ($I_{acc\_th}$)}\label{sec:calib}}
\\ \indent
$I_{acc\_th}$ is calibrated once per model/dataset by searching the prefix of cores 
sorted in ascending $I_{acc}$. 
The smallest prefix whose cumulative importance reduces substitute accuracy to within 
$A_{\mathrm{BB}}+\delta$ is identified, and its cumulative $I_{acc}$ is set as the threshold. 
The full procedure is summarized in \refAlgo{alg:th}. 
This one-time calibration aligns the empirical robustness target with a model-intrinsic threshold, 
decoupling subsequent selection from per-layer sizes.

\begin{algorithm}[t]
\footnotesize
\caption{$I_{acc\_th}$ Calibration via binary search over prefixes}\label{alg:th}
\begin{algorithmic}[1]
\STATE \textbf{Input:} Sorted cores $\{G_i\}$ by ascending $I_{acc}$, B-B robustness $A_{\mathrm{BB}}$, tolerance $\delta$
\STATE \textbf{Output:} $I_{acc\_th}$
\STATE $lo \gets 0$, $hi \gets n$
\WHILE{$lo < hi$}
  \STATE $m \gets \lfloor (lo+hi)/2 \rfloor$; encrypt $\{G_1,\dots,G_m\}$
  \STATE Evaluate substitute accuracy $A_{\mathrm{sub}}(m)$ under \refSection{sec:TTD-threat}
  \STATE \textbf{if} $A_{\mathrm{sub}}(m)\!\le\!A_{\mathrm{BB}}\!+\!\delta$ \textbf{then} $hi\!\gets\!m$
  \STATE \textbf{else} $lo\!\gets\!m\!+\!1$
\ENDWHILE
\STATE $I_{acc\_th} \gets \sum_{i=1}^{lo} I_{acc}(G_i)$
\RETURN $I_{acc\_th}$
\end{algorithmic}
\end{algorithm}

\noindent \textit{\textbf{Minimal Encryption Set: Optimization Formulation}\label{sec:opt}}
\vspace{1pt}

Core sizes differ, so the same threshold may lead to different encryption ratios. We therefore minimize total encrypted parameters subject to meeting the calibrated robustness:
\begingroup
\fontsize{8.5pt}{11pt}\selectfont
\begin{align}\label{eqn:problem}
\textbf{Minimize} \quad & C_{enc} = \sum_{l,k} size(G_{l,k}) \cdot x_{l,k}, \quad x_{l,k}\in\{0,1\}, \\[-2pt]\notag
\textbf{subject to} \quad & \sum_{l,k} I_{acc}(G_{l,k}) \cdot x_{l,k} \ \ge\ I_{acc\_th}.
\end{align}
\endgroup
This is a 0--1 knapsack-type problem with ``cost'' $w_i = size(G_i)$ and ``value'' $v_i = I_{acc}(G_i)$. Our goal is to meet a value threshold with minimum cost.

\noindent \textit{\textbf{Solution Method}\label{sec:solve}}
\vspace{1pt}

The minimal encryption set problem in \refEqn{eqn:problem} is solved using a
\emph{Value-DP} formulation. 
\refAlgo{alg:valdp} details the procedure: values $v_i=I_{acc}(G_i)$ are 
scaled and integerized as $\tilde v_i$, and a dynamic program is used to 
minimize the encryption cost subject to $I_{acc\_th}$. 
We define $\tilde V_{\max}=\sum_i \tilde v_i$, and maintain a DP table
\begin{align*}
dp[i][v] ~=~ \min_{\text{subsets of first }i\text{ cores with total value }=v}\ \text{cost}.
\end{align*}
The recurrence is
\begingroup
\fontsize{8.5pt}{11pt}\selectfont
\begin{align}\label{eqn:valdp}
dp[i][v] = \min\big(dp[i-1][v],\ dp[i-1][v-\tilde v_i] + w_i\big),
\end{align}
\endgroup
with $dp[0][0]=0$ and $dp[0][v>0]=+\infty$. 
The optimal cost is
\begin{align*}
C_{enc\_opt} = \min_{v \ge \tilde V_{th}} dp[n][v],
\end{align*}
where $\tilde V_{th}$ is the integerized counterpart of $I_{acc\_th}$. 
The time and space complexity of the algorithm are both $O(n \cdot \tilde V_{\max})$,
as it maintains the full DP table with backtracking. 
A finer scaling yields larger $\tilde V_{\max}$ (higher precision, higher cost), 
while coarser scaling reduces both.

\begin{algorithm}[t]
\footnotesize
\caption{Value-DP for minimal encryption set}\label{alg:valdp}
\begin{algorithmic}[1]
\STATE \textbf{Input:} $\{(w_i,v_i)\}$ with $w_i=size(G_i)$ and $v_i=I_{acc}(G_i)$, threshold $I_{acc\_th}$
\STATE \textbf{Output:} Minimal cost $C_{enc\_opt}$ and selected core set $S$
\STATE Integerize $v_i \!\rightarrow\! \tilde v_i$ by fixed scaling; compute $\tilde V_{\max}=\sum_i \tilde v_i$ and $\tilde V_{th}$
\STATE Initialize $dp[0][0]=0$ and $dp[0][v>0]=+\infty$; \ \texttt{parent}[i][v]$\gets$none
\FOR{$i=1$ to $n$}
  \STATE \textbf{(Init)} For all $v\in[0,\tilde V_{\max}]$, set $dp[i][v]\gets dp[i-1][v]$
  \FOR{$v=\tilde V_{\max}$ down to $\tilde v_i$}
  \STATE \textbf{if} $dp[i\!-\!1][v\!-\!\tilde v_i]\!+\!w_i < dp[i][v]$ \textbf{then} 
  \STATE \hspace{1.1em} $dp[i][v]\!\gets\!dp[i\!-\!1][v\!-\!\tilde v_i]\!+\!w_i$;\; \texttt{parent}[i][v]$\!\gets\!1$
  \ENDFOR
\ENDFOR
\STATE $C_{enc\_opt} \gets \min_{v\ge \tilde V_{th}} dp[n][v]$; \quad $v^\star \gets \arg\min_{v\ge \tilde V_{th}} dp[n][v]$
\STATE \textbf{(Backtrack)} $S\gets\emptyset$, $i\gets n$, $v\gets v^\star$
\WHILE{$i\ge 1$}
  \STATE \textbf{if} \texttt{parent}[i][v] $=1$ \textbf{then} $S\!\gets\!S\!\cup\!\{G_i\}$;\; $v\!\gets\!v-\tilde v_i$
  \STATE $i\gets i-1$
\ENDWHILE
\RETURN $C_{enc\_opt}, S$
\end{algorithmic}
\end{algorithm}

In summary, TT-SEAL (i) quantifies core criticality via a scale-normalized, efficiently estimable $I_{acc}$, (ii) calibrates $I_{acc\_th}$ from the desired robustness target, and (iii) selects a \emph{minimal} encryption set through an optimization framework.
As shown in \refSection{sec:EXP}, TT-SEAL achieves robustness comparable to the black-box baseline with only a fraction of the parameters encrypted.

\section{Experimental Work}\label{sec:EXP}
In this section, we implement and evaluate TT-SEAL on a prototype FPGA-based edge AI processor. 
We first describe the hardware platform and experimental setup, including the RTL-designed processor, FPGA prototyping, and CIFAR-10-based adversarial attack generation. 
We then assess the robustness of TT-SEAL-encrypted models under the TTD-aware threat model, measuring transferability of adversarial examples across architectures and attack types. 
Finally, we quantify the system-level benefits of TT-SEAL by analyzing encryption cost and decryption overhead in end-to-end inference, demonstrating that selective encryption preserves B-B robustness level while enabling real-time execution on edge hardware.

\subsection{Prototype Processor and Evaluation Setup}\label{sec:setup}

\begin{figure}[t]
\centering
\includegraphics[width=\columnwidth]{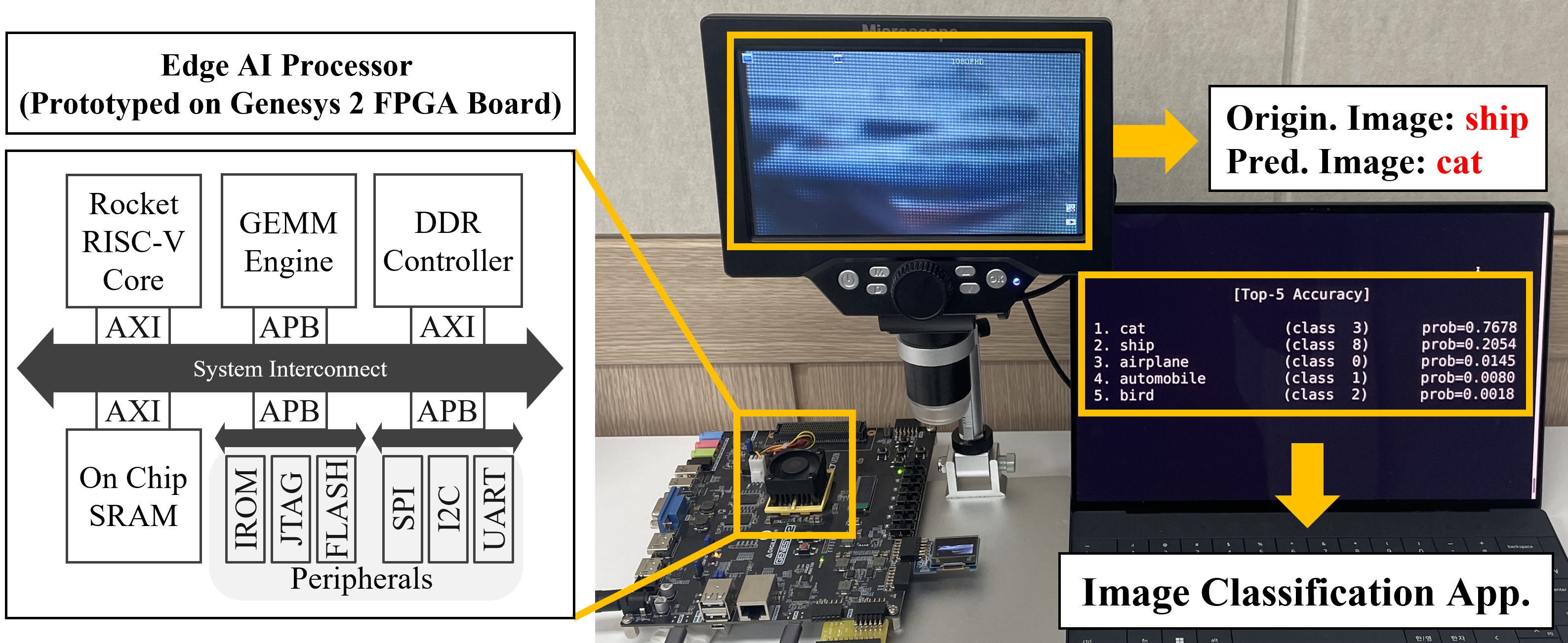}
\vskip -8pt
\caption{Image-classification demo running on the FPGA-prototyped processor, illustrating a transfer-based misclassification: an original \emph{ship} is predicted as a \emph{cat} under the configured threat model.}
\label{fig:fpga}
\vskip -4pt
\end{figure}

All experiments were performed on our RTL-designed and FPGA-prototyped edge AI processor, developed using RISC-V eXpress (RVX), a widely adopted EDA tool for RISC-V processor development~\cite{Han:IoT2021,Park:TCASI24,Lee:IoTJ25,Jeon:DAC2025,Choi:ISLPED2025}.
The processor is based on a single-core RISC-V Rocket microarchitecture~\cite{Rocket}. 
To support the high-bandwidth and low-latency data movement required by AI workloads, 
on-chip SRAMs and peripherals are connected to the core through AXI interfaces, 
and the system interconnect adopts a lightweight ${\mu}\mathrm{NoC}$. 
The processor was prototyped on a Genesys2 Kintex-7 FPGA board~\cite{Genesys2} and operates at 100\,MHz. 
The overall architecture diagram and FPGA prototype photograph are shown in \refFigure{fig:fpga}, 
while \refTable{tab:fpga_resources} summarizes the FPGA resource utilization of the prototype processor.

For application-level evaluation, we used the Darknet framework~\cite{darknet13}, which offers a lightweight C-based implementation of CNNs suitable for FPGA prototyping. We evaluated TTD-compressed variants of ResNet-18, MobileNetV2, and VGG-16. 
To instantiate the threat model, of the CIFAR-10 training images, $45{,}000$ were used to train the original (oracle) model, and the remaining $5{,}000$ served as the attacker's initial seed data for the substitute model. 
The attacker trains the substitute via JBDA: beginning with the $5{,}000$ seeds, iterative dataset augmentation expands the labeled set up to $45{,}000$; each augmentation iteration is followed by $10$ training epochs with learning rate $0.01$ and SGD as the optimizer. 
Adversarial examples are generated using I-FGSM (Iterative-FGSM) based on \refEqn{eqn:FGSM}, with $15$ iterations each; in total, $10{,}000$ adversarial examples are used for evaluation.

\begin{table}[t]
\caption{FPGA resource utilization of the prototype edge AI processor.}
\vskip -6pt
\centering
\renewcommand{\arraystretch}{1.18}
\setlength{\tabcolsep}{2.8pt}
\resizebox{\columnwidth}{!}{
\begin{tabular}{lcc|lcc}
\toprule
\textbf{IP Block} & \textbf{LUTs} & \textbf{FFs} & \textbf{IP Block} & \textbf{LUTs} & \textbf{FFs} \\ 
\hline
Rocket RISC-V Core  & 15,040 & 9,880  & Peripherals (incl. DMA) & 5,043  & 10,370 \\ 
SRAM                & 166    & 323    & System Interconnect     & 9,745  & 17,375 \\ 
DDR Controller      & 7,950  & 7,571  & GEMM Engine             & 84,150 & 32,939 \\ 
\bottomrule
\end{tabular}
}
\label{tab:fpga_resources}
\end{table}

For encryption, we adopt the standard AES scheme. {TT-SEAL} is applied to the convolution layers of each model, while the network input (first convolution layer) and output (final fully connected layer) are {always fully encrypted}. Accordingly, the \emph{white-box} setting means the attacker can access all convolution-layer weights; even in this case, the first and last layers remain encrypted without exception. If the attacker obtains a selectively encrypted model produced by {TT-SEAL}, the exposed parameters are used as-is for learning the substitute, and the encrypted remainder is randomly initialized to construct the substitute model. As a consequence of this setup, \refFigure{fig:fpga} shows a demo in which an image of a ship is misclassified as a cat on the FPGA prototype under our configured threat model.

\subsection{Evaluation}\label{sec:evaluation}

\begin{table*}[!t]
\centering
\caption{Transfer ratios (\%) across ResNet-18, MobileNetV2, and VGG-16 for different perturbation strengths $\boldsymbol{\epsilon}$ and attack types. Results are shown under three encryption levels: W-B (white-box), TH ($\boldsymbol{I}_{\boldsymbol{acc\_th}}$), and B-B (black-box). The comparison highlights how TT-SEAL suppresses transferability to the B-B robustness level.}
\vskip -8pt
\renewcommand{\arraystretch}{1.2}
\setlength{\tabcolsep}{1.5pt}
\resizebox{\textwidth}{!}{%
\begin{tabular}{c|
ccc|ccc|ccc|ccc|
ccc|ccc|ccc|ccc|
ccc|ccc|ccc|ccc}
\toprule
Models
& \multicolumn{12}{c|}{ResNet-18} 
& \multicolumn{12}{c|}{MobileNetV2} 
& \multicolumn{12}{c}{VGG-16} \\ \hline
$\epsilon$
& \multicolumn{3}{c|}{$\epsilon=0/255$} & \multicolumn{3}{c|}{$\epsilon=4/255$} & \multicolumn{3}{c|}{$\epsilon=8/255$} & \multicolumn{3}{c|}{$\epsilon=16/255$}
& \multicolumn{3}{c|}{$\epsilon=0/255$} & \multicolumn{3}{c|}{$\epsilon=4/255$} & \multicolumn{3}{c|}{$\epsilon=8/255$} & \multicolumn{3}{c|}{$\epsilon=16/255$}
& \multicolumn{3}{c|}{$\epsilon=0/255$} & \multicolumn{3}{c|}{$\epsilon=4/255$} & \multicolumn{3}{c|}{$\epsilon=8/255$} & \multicolumn{3}{c}{$\epsilon=16/255$} \\ \hline
\diagbox[width=3.5em,height=1.8em]%
  {\raisebox{-0.3ex}{\scriptsize Att.}}{\raisebox{0.5ex}{\scriptsize $I_{acc}$}}
& W-B & TH & B-B
& W-B & TH & B-B
& W-B & TH & B-B
& W-B & TH & B-B
& W-B & TH & B-B
& W-B & TH & B-B
& W-B & TH & B-B
& W-B & TH & B-B
& W-B & TH & B-B
& W-B & TH & B-B
& W-B & TH & B-B
& W-B & TH & B-B \\ \hline
NT    
&  6.51 &  6.51 &  6.51 & 87.21 &  9.32 &  9.1  & 98.8  & 13.82 & 13.81 & 99.97 & 34.88 & 34.79
&  7.88 &  7.88 &  7.88 & 59.35 & 10.77 & 10.97 & 89.43 & 14.69 & 14.8  & 98.71 & 23.69 & 24.23
&  7.85 &  7.85 &  7.85 & 10.57 &  8.38 &  8.29 & 15.02 &  9.34 &  8.64 & 26.49 & 11.66 & 10.59 \\
RD
&  0.59 &  0.71 &  0.59 & 49.88 &  0.95 &  1.06 & 84.82 &  1.8  &  1.87 & 96.22 &  4.95 &  3.45
&  0.97 &  0.85 &  0.97 & 20.51 &  1.39 &  1.44 & 53.23 &  2.06 &  1.93 & 81.6  &  3.82 &  3.51
&  0.82 &  0.94 &  0.96 &  1.38 &  0.83 &  0.89 &  2.43 &  1.09 &  1.06 &  4.46 &  1.34 &  1.21 \\
SM
&  0.0  &  0.0  &  0.0  & 86.18 &  3.03 &  3.15 & 98.66 &  6.61 &  6.42 & 99.71 & 15.17 & 14.15
&  0.0  &  0.0  &  0.0  & 53.24 &  4.01 &  4.12 & 82.11 &  6.32 &  6.69 & 95.38 & 10.72 & 10.98
&  0.0  &  0.0  &  0.0  &  4.38 &  1.53 &  1.25 &  8.31 &  2.9  &  1.85 & 14.01 &  4.51 &  4.32 \\
LL
&  0.0  &  0.0  &  0.0  & 18.63 &  0.0  &  0.0  & 67.36 &  0.02 &  0.01 & 92.1  &  0.25 &  0.39
&  0.0  &  0.0  &  0.0  &  6.41 &  0.0  &  0.0  & 40.57 &  0.01 &  0.01 & 75.31 &  0.7  &  0.56
&  0.0  &  0.0  &  0.0  &  0.0  &  0.0  &  0.0  &  0.0  &  0.0  &  0.0  &  0.53 &  0.02 &  0.0  \\
\bottomrule
\end{tabular}}
\label{tab:attack_results}
\end{table*}

\subsubsection{Adversarial Robustness}
Under the threat model in \refSection{sec:TTD-threat}, the accuracy of the substitute model that generates adversarial examples is a key determinant of transfer-attack success. 
We evaluate how effectively {TT-SEAL} suppresses the substitute's accuracy and thereby blocks transfer to the oracle $O$ (original model).

Adversarial attacks are categorized as \emph{non-targeted} and \emph{targeted}. 
Non-targeted attacks induce arbitrary misclassification relative to the ground-truth label, whereas targeted attacks force the oracle to predict an attacker-specified target label. 
We define the transferability from the substitute $F$ to the oracle $O$ as
\begin{align}\label{eqn:trans}
T_{F \rightarrow O} =
\begin{cases}
\frac{1}{N} \sum\limits_{i=1}^N 
\big[ O(\mathbf{x}_{adv,i}) \neq y_i \big],
& \textit{(non-target)} \\
\frac{1}{N} \sum\limits_{i=1}^N 
\big[ O(\mathbf{x}_{adv,i}) = y_{t,i} \big],
& \textit{(target)}
\end{cases}
\end{align}
where $[\cdot]$ denotes the Iverson bracket (1 if the condition holds, 0 otherwise), $y_{t,i}$ is the target label for sample $i$, and $N$ is the number of evaluation samples. Larger $T$ indicates stronger transfer (more successful attacks); smaller $T$ indicates transfer is blocked and the oracle maintains correct behavior.

\begin{figure}[t]
\centering
\includegraphics[width=0.97\columnwidth]{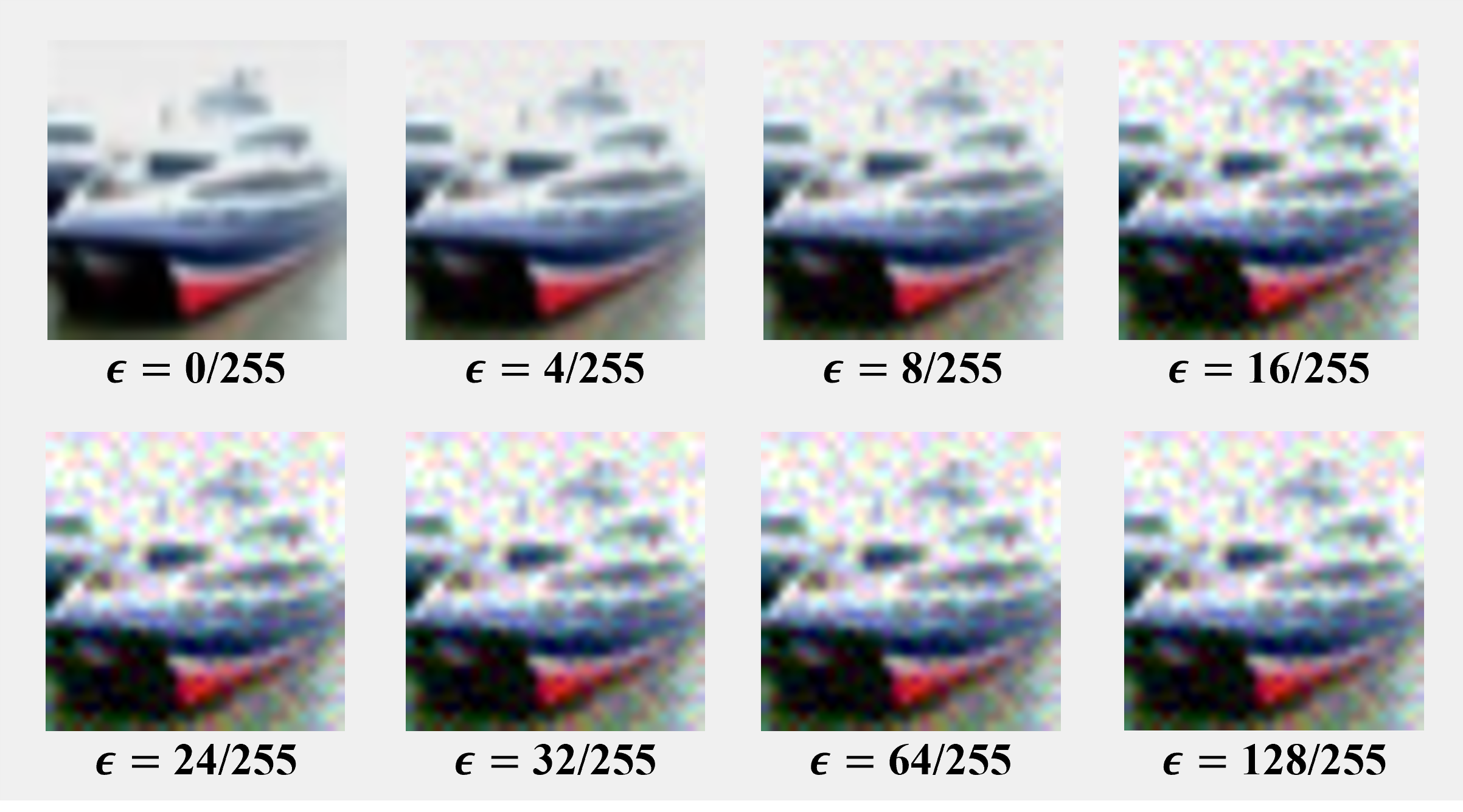}
\vskip -8pt
\caption{Visualization of adversarial examples generated from the substitute of the {TT-SEAL}-encrypted model. As $\boldsymbol{\epsilon}$ increases, stronger perturbations are progressively added to the input image.}
\label{fig:examples}
\vskip -4pt
\end{figure}

We first calibrate the selective-encryption threshold $I_{acc\_th}$ for each model as described in \refFigure{fig:I_acc}; the thresholds are $0.44$ for ResNet-18, $0.79$ for MobileNetV2, and $0.65$ for VGG-16. Using substitutes trained at these $I_{acc\_th}$ points, we generate adversarial examples. \refFigure{fig:examples} visualizes examples for varying $\epsilon$: small $\epsilon$ yields imperceptible perturbations, whereas larger $\epsilon$ produces visibly distorted inputs. 
We evaluate up to $\epsilon=16/255$, which is widely regarded as the practical upper bound on CIFAR-10, beyond which perturbations become visually obvious.

\refTable{tab:attack_results} reports the transfer ratio (\%) computed from $T$ across $\epsilon$, attack types, and encryption levels: \textit{White-box} (W-B), threshold ($I_{acc\_th}$; TH), and \textit{Black-box} (B-B). The attack types include non-target (NT), random target (RD), second most likely target (SM), and least-likely target (LL).

First, the W-B transfer ratios show that as $\epsilon$ increases, transfer rises steeply beyond $90\%$; for ResNet-18, NT and SM attacks approach $100\%$. 
This confirms that, under the worst-case assumption in which the attacker knows internal parameters and structure, our threat model and setup yield strong attacks even on TTD-compressed models.

In contrast, {TT-SEAL} at the TH level effectively suppresses transfer to the B-B level even at large perturbations such as $\epsilon=16/255$. For example, in the ResNet-18 model at $\epsilon=8/255$ with SM attacks, the absolute W-B vs.\ TH gap reaches $92.1\%$. Across cases where the W-B transfer exceeds $50\%$, TH achieves on average a relative reduction exceeding $90\%$, demonstrating that the proposed method remains effective even under severe conditions.

Moreover, the effect is consistent across models and attack types. For ResNet-18, MobileNetV2, and VGG-16 alike, TH results closely match B-B, indicating broad applicability irrespective of network structure. Notably, TH suppresses transfer to the B-B level not only for RD and LL (typically easier to defend) but also for higher-success NT and SM attacks. The absolute TH–B-B difference is at most $1.5\%$ across all cases and averages $0.2\%$, showing that {TT-SEAL}-based selective encryption maintains adversarial robustness at the B-B level across diverse conditions.

\begin{figure}[t]
\centering
\includegraphics[width=0.83\columnwidth]{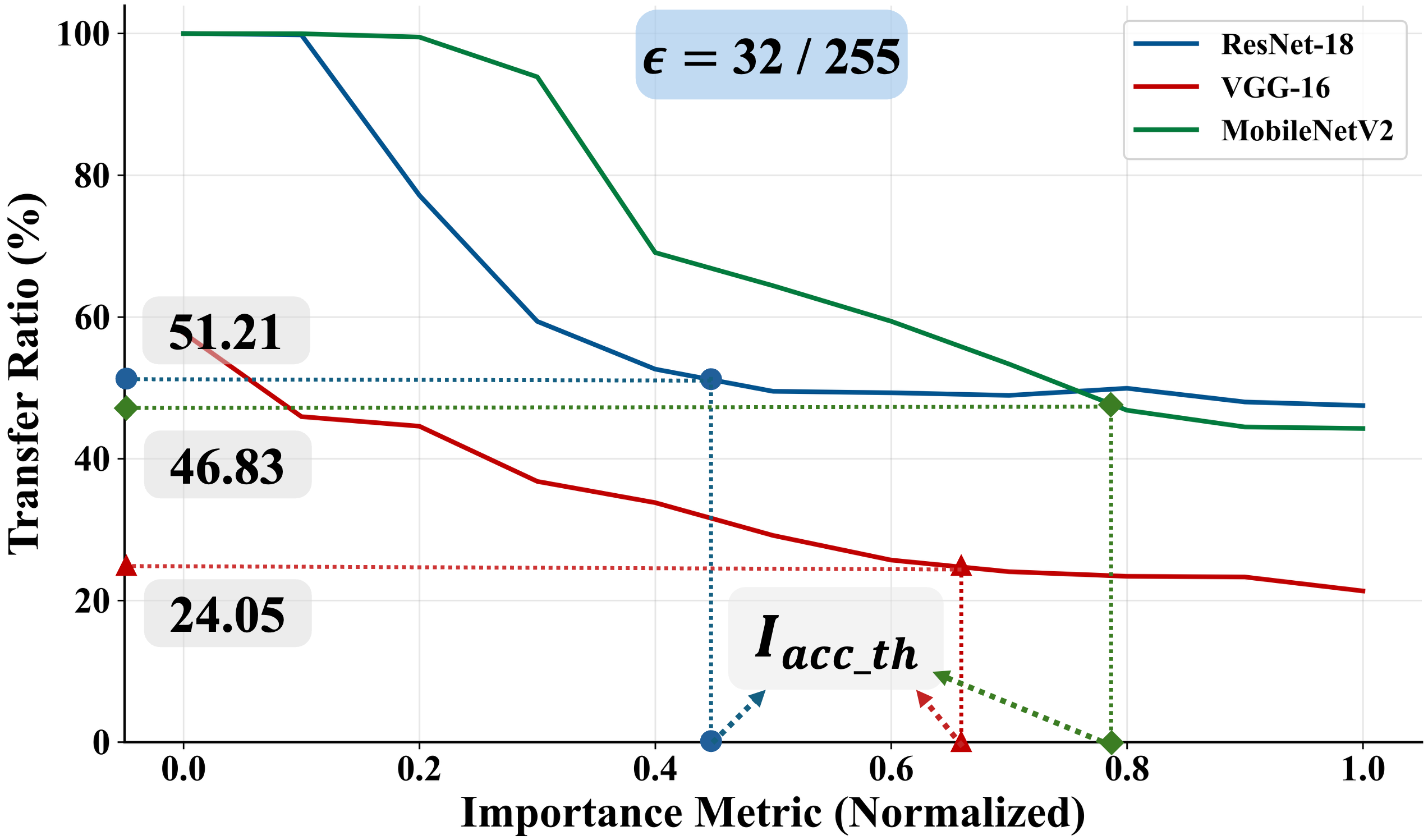}
\vskip -8pt
\caption{Effect of selective encryption under non-targeted attacks ($\boldsymbol{\epsilon=32/255}$). 
Across ResNet-18, VGG-16, and MobileNetV2, the transfer ratio remains suppressed until $\boldsymbol{I}_{\boldsymbol{acc}}$ falls below $\boldsymbol{I}_{\boldsymbol{acc\_th}}$, demonstrating its effectiveness as a robustness threshold.}
\label{fig:adversarial}
\vskip -4pt
\end{figure}

To further stress the setting and to verify that $I_{acc\_th}$ acts as a threshold for selective encryption, we analyze non-target transfer at $\epsilon=32/255$ while sweeping $I_{acc}$, as shown in \refFigure{fig:adversarial}. 
Although the magnitude of change varies by model, all show a sharp increase in transfer once the selective-encryption ratio drops below $I_{acc\_th}$. This indicates that minimal selective encryption at or above $I_{acc\_th}$ is sufficient to suppress transfer, providing adversarial robustness that converges to the B-B level.



\begin{figure*}[t]
\centering
\includegraphics[width=1.99\columnwidth]{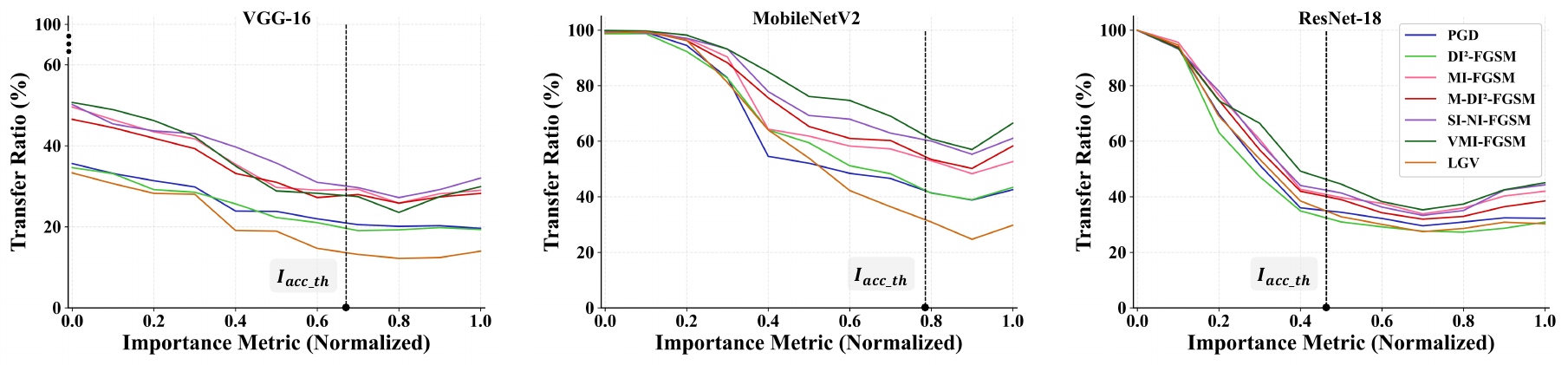}
\vskip -8pt
\caption{Transfer ratios across models for diverse adversarial example generation methods under the non-targeted attack with $\boldsymbol{\epsilon = 16/255}$.}
\label{fig:SOTA}
\end{figure*}

To verify that TT-SEAL generalizes beyond I-FGSM, we evaluate seven transfer-attack 
generators compatible with our threat model and operating on the substitute model: 
PGD~\cite{madry2018towards}, DI$^{2}$-FGSM~\cite{xie2019improving}, MI-FGSM~\cite{dong2018boosting}, 
M-DI$^{2}$-FGSM~\cite{xie2019improving}, SI-NI-FGSM~\cite{lin2019nesterov}, 
VMI-FGSM~\cite{wang2021enhancing}, and LGV~\cite{gubri2022lgv}. 
\refFigure{fig:SOTA} summarizes non-targeted transfer ratios at $\epsilon=16/255$ for 
ResNet-18, MobileNetV2, and VGG-16. 
Despite differences in absolute transferability across generators, all attacks are 
suppressed to the B-B level at the importance threshold $I_{\text{acc\_th}}$, 
with a sharp rise in transferability observed only below this point. 
These results indicate that $I_{\text{acc\_th}}$ generalizes across attack generators 
and architectures, further validating the robustness of TT-SEAL.

\subsubsection{TT-SEAL Performance Analysis via Optimization}
We now quantify performance gains obtained at the optimized selective-encryption ratio found by TT-SEAL. \refTable{tab:performance} summarizes, for each model (ResNet-18, MobileNetV2, VGG-16), the number of encrypted parameters and the decryption-time ratios at $I_{acc}\in\{0.2,0.4,0.6,0.8\}$ and at $I_{acc\_th}$. 
Here, $t_{TT-SEAL}$ and $t_{BB}$ denote the decryption time with {TT-SEAL} (optimized selective encryption) and with full decryption, respectively; $t_{Inf}$ denotes the end-to-end inference time including AES decryption, TTD decoding, and the forward pass.

Because $I_{acc}$ is not directly determined by core size, a core’s importance does not necessarily correlate with its parameter count. 
Empirically, applying TT-SEAL's optimization tends to produce a generally monotonic trend: as $I_{acc}$ increases and the search becomes more selective, the number of encrypted parameters increases gradually. 
This indicates that $I_{acc}$-based selection can modulate encryption cost without relying on raw core size.

Across models, the total parameter counts are $1.83$M (ResNet-18), $447$k (MobileNetV2), and $1.51$M (VGG-16). At $I_{acc\_th}$, the encrypted parameters are about $89$k ($4.89\%$), $71$k ($15.92\%$), and $98$k ($6.46\%$), respectively. ResNet-18 and VGG-16 require only $\sim$5–6\% to secure robustness, whereas MobileNetV2—due to its smaller parameter budget and deeper, more fragmented TT-cores—distributes importance more uniformly and therefore selects more cores, yet still needs only about one-sixth of parameters to reach B-B level.

In terms of decryption time, reducing the encrypted parameter count shrinks $t_{TT-SEAL}$ to as little as $4.87\%$ of $t_{BB}$. Moreover, the decryption share of B-B end-to-end inference time-$58\%$ (ResNet-18), $41.8\%$ (MobileNetV2), and $39\%$ (VGG-16)-drops to $2.76\%$, $6.15\%$, and $2.42\%$ under TT-SEAL, respectively. By driving AES decryption overhead down to the low single digits, TT-SEAL preserves security while enabling real-time inference on edge devices.

\renewcommand{\arraystretch}{1.2}
\begin{table}[t]
\vskip -8pt
\centering
\caption{Selective-encryption parameters and decryption times for each model under $\boldsymbol{I}_{\boldsymbol{acc}}$ intervals and at $\boldsymbol{I}_{\boldsymbol{acc\_th}}$.}
\vskip -8pt
\resizebox{1.0\columnwidth}{!}{
\begin{tabular}{c|c|c|c|c|c|c}
\hline
\multicolumn{2}{c|}{$I_{acc}$} & 0.2 & 0.4 & 0.6 & 0.8 & $I_{acc\_th}$ \\
\hline
\multirowcell{3}[-2.5ex][c]{\textbf{ResNet-18}} 
  & Encrypted Parameters      			 & 21,028   & 72,208    & 134,604   & 465,068   & 89,685 \\
  & Encryption Ratio (\%)     		    & 1.14     & 3.94      & 7.35      & 25.41     & 4.89   \\ \cline{2-7}
  & $t_{TT-SEAL}$ / $t_{BB}$ (\%)   	       & 1.13     & 4.16      & 7.23      & 25.7      & 4.87   \\
  & $t_{TT-SEAL}$ / $t_{Inf}$ (\%) 		    & 0.66     & 2.37      & 4.04      & 13.02     & 2.76   \\
\hline
\multirowcell{3}[-2.5ex][c]{\textbf{MobileNetV2}} 
  & Encrypted Parameters      			 & 15,317   & 19,143    & 29,794    & 74,706   & 71,268 \\
  & Encryption Ratio (\%)     		    & 3.42     & 4.29      & 6.66      & 16.7     & 15.92   \\ \cline{2-7}
  & $t_{TT-SEAL}$ / $t_{BB}$ (\%)   	       & 3.39     & 4.52      & 6.62      & 16.13    & 15.66   \\
  & $t_{TT-SEAL}$ / $t_{Inf}$ (\%) 		    & 1.39     & 1.85      & 2.70      & 6.32     & 6.15    \\
\hline
\multirowcell{3}[-2.5ex][c]{\textbf{VGG-16}} 
  & Encrypted Parameters      			 & 9,699    & 29,364    & 65,052    & 297,705   & 97,513 \\
  & Encryption Ratio (\%)     		    & 0.64     & 1.94      & 4.31      & 19.72     & 6.46   \\ \cline{2-7}
  & $t_{TT-SEAL}$ / $t_{BB}$ (\%)   	       & 0.63     & 2.05      & 4.28      & 19.06     & 6.35   \\
  & $t_{TT-SEAL}$ / $t_{Inf}$ (\%) 		    & 0.25     & 0.79      & 1.64      & 6.92      & 2.42   \\
\hline
\end{tabular}
}
\label{tab:performance}
\end{table}
\renewcommand{\arraystretch}{1.0}

\section{Conclusion}
This paper presented \emph{TT-SEAL}, a TTD-aware selective encryption framework for secure and efficient edge AI.
TT-SEAL exploits the structure of TT-format weights to protect only a minimal set of critical cores: it scores cores with a scale-normalized importance metric that aggregates loss and output sensitivities, calibrates a data-driven threshold aligned with a robustness target, and selects a minimum-cost set of TT cores to encrypt through an optimization procedure.
Under a compression-aware threat model and on an FPGA-prototyped edge AI processor, TT-SEAL suppresses adversarial transfer while encrypting as little as $4.89\%$ of parameters (ResNet-18; $6.46\%$ for VGG-16 and $15.92\%$ for MobileNetV2).
Consequently, the share of AES decryption in end-to-end inference drops from double digits to the low single digits across our models (e.g., $58\%\!\rightarrow\!2.76\%$ on ResNet-18; $41.8\%\!\rightarrow\!6.15\%$ on MobileNetV2; $39\%\!\rightarrow\!2.42\%$ on VGG-16), confirming that TT-SEAL preserves security while satisfying edge latency constraints.
Taken together, these results demonstrate that compression and security can be jointly achieved: by aligning protection with the TT structure, TT-SEAL delivers robustness comparable to full encryption with markedly lower decryption overhead, enabling practical deployment of TTD-compressed models at the edge.

\bibliographystyle{unsrt}
\bibliography{reference}
\end{document}